# Strictly Proper Mechanisms with Cooperating Players

**SangIn Chun and Ross D. Shachter**
Department of Management Science and Engineering
Stanford University
Stanford, CA 94305, USA
*sangin_chun@stanford.edu, shachter@stanford.edu*

## Abstract

Prediction markets provide an efficient means to assess uncertain quantities from forecasters. Traditional and competitive strictly proper scoring rules have been shown to incentivize players to provide truthful probabilistic forecasts. However, we show that when those players can cooperate, these mechanisms can instead discourage them from reporting what they really believe. When players with different beliefs are able to cooperate and form a coalition, these mechanisms admit arbitrage and there is a report that will always pay coalition members more than their truthful forecasts. If the coalition were created by an intermediary, such as a web portal, the intermediary would be guaranteed a profit.

## 1 Introduction

Decision makers need to understand and evaluate complex decision problems with considerable uncertainties. Those uncertainties are integrated into decision models as forecasts in the form of probability distributions, and the analysis is often quite sensitive to the distributions used. Statistical analysis can support this process when there is sufficient relevant historical data. However, in many cases, decision makers must rely instead on the judgment of experts to obtain useful probabilistic forecasts.

In many applications of decision analysis, multiple individual judgments are engaged. Even when the decision maker has her own belief about the uncertainties, she can benefit by integrating the judgments of others. Incorporating diverse views from heterogeneous disciplines has been shown to improve assement accuracy (Clemen 1989, Hoffmann et al. 2007). Several approaches have been developed in the literature to find optimal combinations of multiple individuals' judgements (Clemen and Winkler 1999, Stone 1961, Morris 1977).

A strictly proper scoring rule measures the quality of a probabilistic forecast based on the observed outcome (Winkler 1996). It can be used, ex post, to evaluate the quality of forecasters (Brier 1950), or, ex ante, to encourage forecasters to articulate their best possible report (Savage 1971). In either case, a scoring rule captures the accuracy, calibration, knowledge, and expertise in assessment, encouraging a forecaster to be careful and honest.

A decision maker who needs more knowledge about a particular uncertainty can retain a panel of experts. Each expert reports his forecast to her, and his payment can depend on the observed state, his report, and the other reports, too. Although her contract with the experts can provide a risk-free subsidy, it can also have incentives for superior forecasts (Clemen 2002).

In recent years, prediction markets have been created and several market designs have been developed (Pennock 2004, Peters et al. 2006, Agrawal et al. 2009). Prediction markets can elicit forecasts by aggregating the reports from many anonymous players integrating their combined information (Berg and Rietz 2003). Some prominent prediction markets are Iowa Electronic Markets, which focuses on political election, Hollywood Stock Exchange, which focuses on movies, NewsFutures.com, which covers politics, finance, sports and current international events. It has been shown that forecasts from public prediction markets strongly correlate with observed outcomes (Pennock et al. 2001, Servan-Schreiber et al. 2004).

The most popular type of contract in prediction markets is winner-take-all, where a player is paid if and only if a specific event occurs (Wolfers and Zitzewitz 2004). A winner-take-all market does not induce truthful individual reports. Depending on the assumptions made, the equilibrium price of a prediction market is either a particular quantile of the budget-weighted distribution of players' beliefs (Manski 2006) or the mean of those beliefs (Wolfers and Zitzewitz 2005).

Strictly proper scoring rules have also been applied to groups of forecasters, where the payment depends on the forecaster's score relative to others. Hanson (2003) introduced market scoring rules where the market subsidizes the process. Kilgour and Gerchak (2004) developed competitive scoring rules which are self-financing regardless of the number of players, their forecasts, or the outcome. Johnstone (2007) elaborated the relationship between the Kilgour-Gerchak logarithmic scoring rules and the rewards from Kelly betting competition. Lambert et al. (2008) proposed weighted-score mechanisms, a modified form of Kilgour-Gerchak competitive scoring rules, which ensure budget balance and non-negative payments.

In this paper we explore payments for probabilistic forecasts, either from a panel of experts or in a prediction market. We consider strictly proper mechanisms, both traditional and competitive scoring rules, which have been shown to encourage forecasters to report their true beliefs when acting independently. We show that when forecasters with different beliefs are able to cooperate and form a coalition, these mechanisms admit arbitrage. A similar result for concave traditional scoring rules was suggested by French (1985). These mechanisms actually discourage the coalition members from reporting their true beliefs because they are guaranteed to receive more when they coordinate their reports. If the coalition were created by an intermediary, such as a web portal, without the knowledge of the players, the intermediary would be guaranteed a profit.

It should not be surprising that mechanisms that encourage individuals acting alone to report honestly, do otherwise when players can cooperate. This is similar to results with auctions. Although second-price auctions are incentive-compatible for individuals, cooperation among bidders is a dominant strategy that affects the reserve price for the auction (Graham and Marshall 1987). Auctions with intermediaries are similary influenced by cooperative behavior (Feldman et al. 2010). Such cooperation is difficult to detect and ambiguous to prove (McAfee and McMillan 1992).

In section 2 of this paper, we present the fundamental notation and earlier results for traditional and competitive scoring rules. We build on them to show how strictly proper mechanisms admit arbitrage under cooperation for traditional and competitive scoring rules, in sections 3 and 4, respectively. Finally, we conclude and present some future research in section 5.

## 2 Traditional and Competitive Scoring Rules

A decision maker is concerned about a random variable $E$ with $m$ mutually exclusive and collectively exhaustive states given by $\Omega = \{E_1, E_2, ..., E_m\}$. There are $n$ *forecasters* or *players*, indexed by $\mathcal{N} = \{1, 2, ..., n\}$, who report probabilistic forecasts for $E$. The players are assumed to be risk neutral and their forecasts are in the simplex, denoted by $\Delta_m = \left\{ \mathbf{r} \in \mathbb{R}^m : \sum_{j=1}^m r_j = 1, r_j \geq 0 \right\}$. We assume that each player $i \in \mathcal{N}$ believes that the event will occur according with probability $\mathbf{p}_i = (p_{i1}, ..., p_{im}) \in \Delta_m$ but reports $\mathbf{r}_i \in \Delta_m$ to the decision maker, where $\mathbf{r}_i$ is not necessarily equal to $\mathbf{p}_i$. We say that the players in $\mathcal{C} \subseteq \mathcal{N}$ *agree* if $\mathbf{p}_1 = \mathbf{p}_i$ for all $i \in \mathcal{C}$, and that otherwise they *disagree*. The decision maker pays the players according to a *contract function*, denoted by $\Pi : (\Delta_m)^n \times \Omega \to \mathbb{R}^n$ based on the reports of all players and the observed state, so player $i$ receives $\Pi_i(\mathbf{r}_1, ..., \mathbf{r}_n, E_j)$.

A traditional *scoring rule*, $S : \Delta_m \times \Omega \to \mathbb{R}$, assigns a real-valued score based solely on the player's report and the observed state. If the decision maker pays player $i$

$$\Pi_i(\mathbf{r}_1, ..., \mathbf{r}_n, E_j) = w_i S(\mathbf{r}_i, E_j),$$

when $E_j \in \Omega$ is observed, where *weight* $w_i > 0$, $\Pi$ is said to be a *contract function for scoring rule $S$*. A scoring rule $S$ is said to be *strictly proper* for player $i$ if

$$\sum_{j=1}^m p_{ij} S(\mathbf{p}_i, E_j) > \sum_{j=1}^m p_{ij} S(\mathbf{r}_i, E_j)$$

whenever $\mathbf{p}_i \neq \mathbf{r}_i$, for all $\mathbf{p}_i, \mathbf{r}_i \in \Delta_m$. The contract function $\Pi$ is *strictly proper* if

$$\sum_{j=1}^m p_{ij} \Pi_i(\mathbf{r}_1, ..., \mathbf{p}_i, ..., \mathbf{r}_n, E_j) > \sum_{j=1}^m p_{ij} \Pi_i(\mathbf{r}_1, ..., \mathbf{r}_n, E_j)$$

whenever $\mathbf{p}_i \neq \mathbf{r}_i$, for all $i$ and all $\mathbf{p}_i, \mathbf{r}_1, ..., \mathbf{r}_n \in \Delta_m$. Thus, under a a strictly proper mechanism, a risk neutral player should report his true belief in order to maximize his expected score. Savage (1971) showed that scoring rule $S$ is strictly proper if and only if

$$S(\mathbf{r}_i, E_j) = G(\mathbf{r}_i) - \nabla G(\mathbf{r}_i)\mathbf{r}_i + \nabla_j G(\mathbf{r}_i) \quad (1)$$

for $j = 1, ..., m$, where $G : \Delta_m \to \mathbb{R}$ is a continuously differentiable strictly convex function.

The three strictly proper scoring rules most commonly used in the literature are the quadratic (or Brier), logarithmic, and spherical rules: (Winkler 1996, Jose 2008):

**Quadratic:** $S(\mathbf{r}_i, E_j) = a_j + b(2r_{ij} - \|\mathbf{r}_i\|^2)$,

**Logarithmic:** $S(\mathbf{r}_i, E_j) = a_j + b \log r_{ij}$, and

**Spherical:** $S(\mathbf{r}_i, E_j) = a_j + \dfrac{br_{ij}}{\|\mathbf{r}_i\|}$,

for any $a_j$ and any $b > 0$, where $\|\mathbf{r}_i\| = \left(\sum r_{ij}^2\right)^{1/2}$, the 2-norm of the vector $\mathbf{r}_i$.

Of the three most common scoring rules, the quadratic and spherical are bounded, but the logarithmic has no lower bound, which makes it less attractive as a contract payment. Therefore we propose a new strictly proper scoring rule, a *generalized logarithmic score* given by

**Generalized Logarithmic:**

$$S(\mathbf{r}_i, E_j) = a_j + b\log(r_{ij} + l) + bl \sum_{k=1}^{m} \log(r_{ik} + l), \quad (2)$$

which is equal to the logarithmic score when $l = 0$ and bounded when $l > 0$ for any $a_j$ and any $b > 0$.

Kilgour and Gerchak (2004) construct a self-financing *competitive scoring rule*, where the total of the contract functions for all players is zero. Their contract function assigns a score based on the relative quality of each forecast compared to the other forecasts. If $E_j \in \Omega$ is observed, player $i$ receives $\Pi_i(\mathbf{r}_1, ..., \mathbf{r}_n, E_j)$ determined by

$$\Pi_i(\mathbf{r}_1, ..., \mathbf{r}_n, E_j) = S(\mathbf{r}_i, E_j) - \frac{1}{n}\sum_{k \in \mathcal{N}} S(\mathbf{r}_k, E_j) \quad (3)$$

Lambert et al. (2008) propose a similar *weighted-score mechanism*, denoted by

$$\Pi_i(\mathbf{r}_1, ..., \mathbf{r}_n, E_j) = w_i S(\mathbf{r}_i, E_j) - \frac{w_i}{w_\mathcal{N}} \sum_{k \in \mathcal{N}} w_k S(\mathbf{r}_k, E_j) \quad (4)$$

where weight $w_i > 0$ can now be interpreted as the *wager* chosen by player $i$ and $w_\mathcal{C} = \sum_{k \in \mathcal{C}} w_k$ for any $\mathcal{C} \subseteq \mathcal{N}$. They require that scoring rule $S$ be bounded, mapping into $[0, 1]$, to ensure that player $i$ never loses more than his wager $w_i$. If a scoring rule $S$ is strictly proper, then so are the corresponding Kilgour-Gerchak rule and weighted-score mechanisms (Kilgour and Gerchak, 2004; Lambert et al. 2008).

Hanson's *market scoring rule* (2003) is also strictly proper. Player $i$ reports his forecast and is paid based on that report and the report from the player who preceded him, $i - 1$, according to the following contract rule:

$$\Pi_i(\mathbf{r}_1, ..., \mathbf{r}_n, E_j) = S(\mathbf{r}_i, E_j) - S(\mathbf{r}_{i-1}, E_j). \quad (5)$$

Although these mechanisms are all strictly proper in the sense that each player, acting independently, should report his true beliefs in order to maximize his expected score, we will show that is not the case when he can cooperate with other players.

## 3 Arbitrage with Traditional Scoring Rules

Given a contract function for a strictly proper scoring rule, we show how players should behave when they are allowed to cooperate. Suppose that two or more players form a *coalition*, $\mathcal{C} \subseteq \mathcal{N}$ of size $c = |\mathcal{C}| \geq 2$. We show that in a variety of circumstances, even under a strictly proper mechanism, coalition members who disagree about the forecasts will all be better off coordinating their reports than reporting truthfully.

We say that contract $\Pi$ *admits arbitrage* for coalition $\mathcal{C}$ if, for any $\mathbf{p}_i \in \Delta_m$ for all $i \in \mathcal{C}$ where the players in $\mathcal{C}$ disagree, there are $\mathbf{r}_i \in \Delta_m$ for all $i \in \mathcal{C}$ such that

$$\sum_{i \in \mathcal{C}} \Pi_i(\mathbf{p}_1, ..., \mathbf{p}_n, E_j) < \sum_{i \in \mathcal{C}} \Pi_i(\mathbf{r}_1, ..., \mathbf{r}_n, E_j)$$

for all $j$ and all $\mathbf{p}_k = \mathbf{r}_k \in \Delta_m$ for all $k \notin \mathcal{C}$. We compare reports from coalition members with their true beliefs given any possible reports from those outside the coalition.

The contract $\Pi$ admits arbitrage *with identical reports* if $\mathbf{q} = \mathbf{r}_i$ for all $i \in \mathcal{C}$ for some $\mathbf{q}$.

If a contract admits arbitrage for coalition $\mathcal{C}$, then whenever coalition members disagree about $E$ the total payments to the coalition members will always be greater *ex post* reporting $\mathbf{r}_i$ rather than $\mathbf{p}_i$, regardless which outcome is observed and what the other players report. We call the additional payment to the coalition members the *surplus from cooperation*.

### 3.1 Scoring Rules for Binary Events

We consider a binary event sample space $\Omega = \{E_1, E_2\}$, so that each there is a single parameter $r_i$ for player $i$ to report, $\mathbf{r}_i = (r_i, 1 - r_i)$ and the player's belief about $E_1$ is characterized by $p_i$, $\mathbf{p}_i = (p_i, 1 - p_i)$. A traditional strictly proper score $S$ for a binary event can be generally represented by

$$S(r_i, E_1) = G(r_i) + (1 - r_i)G'(r_i),$$
$$S(r_i, E_2) = G(r_i) - r_i G'(r_i),$$

where $r_i$ is the reported probability report for $E_1$ and $G$ is a continuously differentiable strictly convex function (Savage 1971).

**Theorem 1** (Arbitrage with Binary $E$). *Given binary $E$, any contract function for a strictly proper scoring*

rule admits arbitrage with identical reports. Whenever coalition members maximizing their payments disagree about E they should not report their true beliefs.

*Proof.* We will show that the contract function $\Pi_i(r_1,...,r_n,E_j) = S(r_i,E_j)$ admits arbitrage with identical reports $q$ for any coalition $\mathcal{C}$ with $c \geq 2$, where $q$ satisfies

$$G'(q) = \sum_{i \in \mathcal{C}} \frac{w_i}{w_\mathcal{C}} G'(p_i). \tag{6}$$

$G$ strictly convex implies that $G'$ is strictly increasing. Therefore, by the intermediate-value-theorem there exists a unique such $q$, $\min_{i \in \mathcal{C}} p_i < q < \max_{i \in \mathcal{C}} p_i$.

Given such a $q$, there is the same surplus from cooperation no matter which state of $E$ is observed, and because $G$ is strictly convex, the surplus is strictly positive when the coalition members disagree,

$$\sum_{i \in \mathcal{C}} w_i S(q, E_1)$$
$$= \sum_{i \in \mathcal{C}} w_i \left( G(q) + (1-q)G'(q) \right)$$
$$= \sum_{i \in \mathcal{C}} w_i \left( G(q) + (1-q)G'(p_i) \right)$$
$$> \sum_{i \in \mathcal{C}} w_i \left( G(p_i) + (q-p_i)G'(p_i) + (1-q)G'(p_i) \right)$$
$$= \sum_{i \in \mathcal{C}} w_i \left( G(p_i) + (1-p_i)G'(p_i) \right)$$
$$= \sum_{i \in \mathcal{C}} w_i S(p_i, E_1),$$

and

$$\sum_{i \in \mathcal{C}} w_i S(q, E_2) - \sum_{i \in \mathcal{C}} w_i S(p_i, E_2)$$
$$= \sum_{i \in \mathcal{C}} w_i \left( G(q) - qG'(q) \right)$$
$$\quad - \sum_{i \in \mathcal{C}} w_i \left( G(p_i) - p_i G'(p_i) \right)$$
$$= \sum_{i \in \mathcal{C}} w_i S(q, E_1) - \sum_{i \in \mathcal{C}} w_i S(p_i, E_1) > 0.$$

□

### 3.2 Strictly Concave Scoring Rules

In this section, we consider contract functions for strictly concave scoring rules when $E$ has $m \geq 2$ possible states. These include quadratic and generalized logarithmic scoring rules, but spherical scoring rules are not concave. We show that strictly concave scoring rules admit arbitrage for all coalitions, as suggested by French (1985).

**Theorem 2** (Arbitrage with Strictly Concave Scoring Rule). *Any contract function for a strictly concave scoring rule admits arbitrage with identical reports. Whenever coalition members maximizing their payments disagree about E they should not report their true beliefs.*

*Proof.* Consider contract function $\Pi_i(\mathbf{r}_1,...,\mathbf{r}_n,E_j) = w_i S(\mathbf{r}_i, E_j)$ and let the identical reports be

$$\mathbf{q} = \sum_{i \in \mathcal{C}} \frac{w_i}{w_\mathcal{C}} \mathbf{p}_i \in \Delta_m.$$

Whenever the members disagree, the strict concavity of $S$ with respect to $\mathbf{r}_i$ implies that for any $E_j$ the surplus from cooperation is positive,

$$\sum_{i \in \mathcal{C}} w_i S(\mathbf{q}, E_j) - \sum_{i \in \mathcal{C}} w_i S(\mathbf{p}_i, E_j) > 0$$

.

□

### 3.3 Examples of Traditonal Strictly Proper Scoring Rules

The three most commonly used strictly proper scoring rules are the quadratic, logarithmic and spherical rules. The contract function for these scoring rules admits arbitrage with identical reports for any coalitions. Under all three, when players are able to cooperate they can be discouraged from reporting their true beliefs. The results of this section are summarized in Table 1.

**Example 1** (Quadratic Scoring Rule). The quadratic scoring rule for event $E$ is $S(\mathbf{r}_i, E_j) = a_j + b(2r_{ij} - \|\mathbf{r}_i\|^2)$. It is strictly concave with respect to $\mathbf{r}_i$, so it admits arbitrage with identical reports

$$\mathbf{q} = \sum_{i \in \mathcal{C}} \frac{w_i}{w_\mathcal{C}} \mathbf{p}_i \in \Delta_m.$$

In this case, the surplus from cooperation is the same for all outcome states $E_j$,

$$\sum_{i \in \mathcal{C}} w_i S(\mathbf{q}, E_j) - \sum_{i \in \mathcal{C}} w_i S(\mathbf{p}_i, E_j)$$
$$= b \sum_{i \in \mathcal{C}} w_i \left[ \|\mathbf{p}_i\|^2 - \|\mathbf{q}\|^2 \right]$$
$$= b \sum_{i \in \mathcal{C}} w_i \|\mathbf{p}_i - \mathbf{q}\|^2.$$

This surplus comes from the total variation among the coalition members' probabilities, a second moment analogous to variance. It is positive unless they agree on $E$ and it increases when there is more disagreement among the coalition members. For example, if there were two members who disagreed about a binary event, believing probabilities 0.2 and 0.8 for $E_1$, their surplus from cooperating and reporting 0.5 instead of their true beliefs would be nine times the surplus they would receive if they had believed 0.4 and 0.6 and reported 0.5. Their disagreement is three times larger and the surplus is therefore $3^2$ larger.

**Example 2** (Generalized Logarithmic Scoring Rule). The generalized logarithmic scoring rule for event $E$ is $S(\mathbf{r}_i, E_j) = a_j + b\log(r_{ij} + l) + bl \sum_{k=1}^{m} \log(r_{ik} + l)$, where $l = 0$ corresponds to the traditional logarithmic scoring rule. Because it is concave, by Theorem 2 it admits arbitrage with identical reports, equal to the weighted arithmetic mean of the coalition members' forecasts, for any coalition. However, it is more instructive to consider identical reports equal to the normalized weighted geometric mean,

$$q_j = \frac{(1+ml) \prod_{i \in \mathcal{C}} (p_{ij} + l)^{w_i/w_\mathcal{C}}}{\sum_{k=1}^{m} \prod_{i \in \mathcal{C}} (p_{ik} + l)^{w_i/w_\mathcal{C}}} - l. \quad (7)$$

With this $\mathbf{q}$ the surplus from cooperation is

$$bw_\mathcal{C} (1+ml) \log \left[ \frac{1+ml}{\sum_{j=1}^{m} \prod_{i \in \mathcal{C}} (p_{ij} + l)^{w_i/w_\mathcal{C}}} \right] \quad (8)$$

regardless of the observed outcome, and is strictly positive whenever they disagree. By Jensen's inequality, $\prod_{i \in \mathcal{C}} (p_{ij} + l)^{w_i/w_\mathcal{C}} < \sum_{i \in \mathcal{C}} \frac{w_i}{w_\mathcal{C}}(p_{ij} + l)$, unless $p_{ij}$ are the same for all $i \in \mathcal{C}$. Summing up each component, $\sum_{j=1}^{m}(\prod_{i \in \mathcal{C}} (p_{ij} + l))^{w_i/w_\mathcal{C}} < \sum_{j=1}^{m} \sum_{i \in \mathcal{C}} \frac{w_i}{w_\mathcal{C}}(p_{ij} + l) = 1 + ml$, thus the equation (8) is strictly positive.

**Example 3** (Spherical Scoring Rule). The spherical scoring rule for event $E$, $S(\mathbf{r}_i, E_j) = a_j + br_{ij}/\|\mathbf{r}_i\|$, also admits arbitrage with identical reports for any coalition. However, a spherical score is not a concave function, and the arithmetic mean does not always lead to arbitrage. For example, if there were two members who disagreed about a binary event, believing probabilities 0.1 and 0.4 for $E_1$, although they would always be better off both reporting 0.275 than reporting their true beliefs, they are not always better off reporting the arithmetic mean 0.25.

**Theorem 3** (Arbitrage with Spherical Scoring Rule). *Any contract function for a spherical scoring rule admits arbitrage with identical reports. Whenever coalition members maximizing their payments disagree about $E$ they should not report their true beliefs.*

The proof is given in the Appendix. It shows that there is an identical report $\mathbf{q} \in \Delta_m$ with components

$$q_j = \frac{1}{m} + \frac{Y_j - \overline{Y}}{\sqrt{m(1 - \sum_{k=1}^{m}(Y_k - \overline{Y})^2)}} \quad (9)$$

where $Y_j = \sum_{i \in \mathcal{C}} \frac{w_i p_{ij}}{w_\mathcal{C} \|p_i\|}$ and $\overline{Y} = \frac{1}{m} \sum_{j=1}^{m} Y_j$.

The surplus from cooperation is

$$bw_\mathcal{C} \left( \sqrt{\frac{1 - \sum_{j=1}^{m}(Y_j - \overline{Y})^2}{m}} - \overline{Y} \right) \quad (10)$$

regardless of the observed outcome, and is strictly positive whenever they disagree.

## 4 Arbitrage with Competitive Scoring Rules

In Section 3 we showed that contracts for strictly proper scoring rules admit arbitrage when players can cooperate. We build on those results in this section to show how competitive mechanisms, shown to be strictly proper when players act independently, also admit arbitrage when players can cooperate. As a result, such mechanisms can discourage coalition members from reporting truthful forecasts whenever they disagree.

### 4.1 Self-Financed Competitive Mechanisms

We consider a *self-financed competitive mechanism* with contract function given by

$$\Pi_i(\mathbf{r}_1, ..., \mathbf{r}_n, E_j) = w_i S(\mathbf{r}_i, E_j) - \frac{w_i}{w_\mathcal{N}} \sum_{k \in \mathcal{N}} w_k S(\mathbf{r}_k, E_j), \quad (11)$$

where the traditional scoring rule $S$ is strictly proper and the wagers $w_i > 0$ are positive. This mechanism is self-financing, with contract functions summing to zero in all outcome states. It generalizes both the Kilgour-Gerchak and Lambert mechanisms for the purposes of this paper. If all of the wagers are equal, $w_i = w_1$ for all $i \in \mathcal{N}$, this is the Kilgour-Gerchak competitive scoring rule. If the range of the scoring rule is restricted to $[0, 1]$, then it is the Lambert weighted-score mechanism, and each player $i$ never loses more than his wager, $\Pi_i > -w_i$.

**Theorem 4** (Arbitrage with Self-Financed Competitive Mechanism). *Given that the contract function for a strictly proper scoring rule admits arbitrage, any self-financed competitive mechanism using the same scoring rule admits arbitrage for any coalition including at least two but not all of the players. Whenever coalition members maximizing their payments disagree about $E$ they should not report their true beliefs.*

*Proof.* Suppose that the contract function admits arbitrage for coalition $\mathcal{C}$ with each player $i \in \mathcal{C}$ believing $\mathbf{p}_i$ and reporting $\mathbf{r}_i$, while each player $k \notin \mathcal{C}$ reports $\mathbf{r}_k$

Table 1: Examples of identical reports $\mathbf{q}$ for arbitrage in contracts for scoring rules

| Type of scoring rule | Condition for $\mathbf{q}$ |
|---|---|
| Score for Binary Event | $G'(q) = \sum_{i \in \mathcal{C}} \frac{w_i}{w_{\mathcal{C}}} G'(p_i)$ |
| Concave Score | $\mathbf{q} = \sum_{i \in \mathcal{C}} \frac{w_i}{w_{\mathcal{C}}} \mathbf{p}_i$ |
| Quadratic Score | $\mathbf{q} = \sum_{i \in \mathcal{C}} \frac{w_i}{w_{\mathcal{C}}} \mathbf{p}_i$ |
| Generalized Logarithmic Score | $q_j = \frac{(1+ml) \prod_{i \in \mathcal{C}} (p_{ij}+l)^{w_i/w_{\mathcal{C}}}}{\sum_{k=1}^{m} \prod_{i \in \mathcal{C}} (p_{ik}+l)^{w_i/w_{\mathcal{C}}}} - l$ |
| Spherical Score | $q_j = \frac{1}{m} + \frac{Y_j - \overline{Y}}{\sqrt{m(1 - \sum_{k=1}^{m}(Y_k - \overline{Y})^2)}}$ where $Y_j = \sum_{i \in \mathcal{C}} \frac{w_i p_{ij}}{w_{\mathcal{C}} \|p_i\|}, \overline{Y} = \frac{1}{m} \sum_{j=1}^{m} Y_j$ |

and we will let $\mathbf{p}_k = \mathbf{r}_k$. The surplus from cooperation when $E_j$ is observed is

$$\sum_{i \in \mathcal{C}} \Pi_i(\mathbf{r}_1, ..., \mathbf{r}_n, E_j) - \sum_{i \in \mathcal{C}} \Pi_i(\mathbf{p}_1, ..., \mathbf{p}_n, E_j)$$
$$= \sum_{i \in \mathcal{C}} w_i S(\mathbf{r}_i, E_j) - \frac{w_{\mathcal{C}}}{w_{\mathcal{N}}} \sum_{k \in \mathcal{C}} w_k S(\mathbf{r}_k, E_j)$$
$$- \sum_{i \in \mathcal{C}} w_i S(\mathbf{p}_i, E_j) + \frac{w_{\mathcal{C}}}{w_{\mathcal{N}}} \sum_{k \in \mathcal{C}} w_k S(\mathbf{p}_k, E_j)$$
$$= \left(1 - \frac{w_{\mathcal{C}}}{w_{\mathcal{N}}}\right) \sum_{i \in \mathcal{C}} w_i \left[S(\mathbf{r}_i, E_j) - S(\mathbf{p}_i, E_j)\right].$$

Thus the competitive mechanism admits arbitrage for the coalition provided $w_{\mathcal{C}} < w_{\mathcal{N}}$. □

We have shown that there is surplus from cooperation for both the Kilgour-Gerchak competitive scoring rule and the Lambert weighted-score mechanism. In prediction markets, it would be difficult to prevent players from cooperating, and they could in fact do so without even knowing it. An intermediary, such as a web portal, can create a coalition among players, and exploit the surplus from cooperation that comes from bringing together forecasters who disagree. Such an intermediary can provide a convenient service for players who want to enter their forecasts into the prediction market, reimbursing them as if their forecasts were actually entered into the market. However, if the intermediary instead submits identical reports for each of them, or one large wager with that report, he would be guaranteed a profit if any of the players disagree, even if he had no knowledge about the uncertainty. The players might be unaware they had participated in a coalition.

Next we consider how large a coalition should be to maximize the expected surplus.

**Theorem 5** (Optimal Coalition Size). *Given that a large number of players n are participating with equal wagers, their forecasts are believed to be exchangeable, and the contract admits arbitrage with identical reports, the ideal size for a coalition $\mathcal{C}$ in order to maximize the expected surplus from cooperation includes all players under a contract for a strictly proper scoring rule, and half of the wagers, $w_{\mathcal{C}} = \frac{1}{2} w_{\mathcal{N}}$, under a self-financed competitive mechanism.*

*Proof.* For a large enough population, the identical report for members of the coalition $\mathbf{q}$ would not depend on the size of the coalition. Therefore, the expected surplus from cooperation under a contract for a scoring rule when $E_j$ is observed is

$$\sum_{i \in \mathcal{C}} w_i \left[S(\mathbf{q}, E_j) - \int_{\mathbf{p}_i \in \Delta_m} S(\mathbf{p}_i, E_j) dF(\mathbf{p}_i)\right]$$
$$= w_{\mathcal{C}} \left[S(\mathbf{q}, E_j) - \int_{\mathbf{p} \in \Delta_m} S(\mathbf{p}, E_j) dF(\mathbf{p})\right],$$

which is proportional to $w_{\mathcal{C}}$ and it is maximized by including all players. If players are paid by a self financed competitive mechnism, the expected surplus from co-

operation when $E_j$ is observed is

$$\left(1 - \frac{w_\mathcal{C}}{w_\mathcal{N}}\right) \sum_{i \in \mathcal{C}} w_i \left[S(\mathbf{q}, E_j) - \int_{\mathbf{p}_i} S(\mathbf{p}_i, E_j) dF(\mathbf{p}_i)\right]$$
$$= \left(1 - \frac{w_\mathcal{C}}{w_\mathcal{N}}\right) w_\mathcal{C} \left[S(\mathbf{q}, E_j) - \int_{\mathbf{p}} S(\mathbf{p}, E_j) dF(\mathbf{p})\right].$$

This is proportional to $w_\mathcal{C}(w_\mathcal{N} - w_\mathcal{C})$. Therefore, it is maximized when $w_\mathcal{C} = \frac{1}{2} w_\mathcal{N}$. □

In a contract for a scoring rule, the surplus from cooperation is proportional to the size of the coalition, and the surplus is largest with all players cooperating. However, under a competitive mechanism the surplus comes from the players outside the coalition and from the variation among the players within it, and therefore the surplus is largest when half of the wagers are from members, $w_\mathcal{C} = \frac{1}{2} w_\mathcal{N}$. Note that if our objective were to maximize the surplus per member, i.e. the total surplus from cooperation divided by the size of the coalition, we would prefer a smaller coalition.

### 4.2 Market Scoring Rules

We now consider Hanson's market scoring rule, which combines the advantages of scoring rules and standard information markets, and show conditions under which cooperating players maximizing their total payment are discouraged from reporting their true forecasts.

**Theorem 6** (Arbitrage with Market Scoring Rules). *Given that the contract function for a strictly proper scoring rule admits arbitrage, a market scoring rule using the same scoring rule admits arbitrage for any coalition $\mathcal{C} \subseteq \mathcal{N}$ if players outside the coalition report before each coalition member, and those players' reports are independent of the coalition members' reports. Whenever coalition members maximizing their payments disagree about $E$ they should not report their true beliefs.*

*Proof.* Suppose that the contract function admits arbitrage for coalition $\mathcal{C}$ with each player $i \in \mathcal{C}$ believing $\mathbf{p}_i$ and reporting $\mathbf{r}_i$, while each player $k \notin \mathcal{C}$ reports $\mathbf{r}_k$. For any observed outcome $E_j$ the surplus from cooperation is

$$\sum_{i \in \mathcal{C}} [S(\mathbf{r}_i, E_j) - S(\mathbf{r}_{i-1}, E_j)]$$
$$- \sum_{i \in \mathcal{C}} [S(\mathbf{p}_i, E_j) - S(\mathbf{r}_{i-1}, E_j)]$$
$$= \sum_{i \in \mathcal{C}} S(\mathbf{r}_i, E_j) - \sum_{i \in \mathcal{C}} S(\mathbf{p}_i, E_j).$$

□

Although the independence and alternating participation conditions assumed for the theorem are rather strong, they provide a scenario under which the market scoring rule admits arbitrage.

## 5 Conclusions

We have shown that many of the strictly proper mechanisms that have been shown to encourage forecasters to report their true beliefs fail to do so when the forecasters are able to cooperate. When players form a coalition they always receive more by coordinating their reports because the mechanisms admit arbitrage. We have shown this happens with contracts for traditional strictly proper scoring rules, including the quadratic, logarithmic, and spherical scoring rules. We have also shown this arbitrage for competitive mechanisms based on strictly proper scoring rules, such as those proposed by Hanson (2003), Kilgour and Gerchak (2004) and Lambert et al. (2008).

We had been hoping to use these results to develop a mechanism resistant to cooperation. It is still an open question whether there is any strictly proper mechanism that does not admit arbitrage, but it seems unlikely. In prediction markets, it would be difficult to prevent players from cooperating, or to prevent an intermediary, such as a web portal, from exploiting the surplus from cooperation that comes from bringing together forecasters who disagree. Such an intermediary would be guaranteed a profit even if she had no knowledge about the uncertainty.

For a decision maker subsidizing a prediction market to observe players' true forecasts, these results raise questions. They suggest that coalitions obtain their surplus from cooperation by reducing the variation among the reports from members, falsely appearing to reach consensus. Thus, decision makers would lose valuable information about the diversity of opinion among forecasters, and might lose the benefit of their heterogeneous information sources.

Another issue with cooperation is that the surplus from cooperation can increase the cost to the decision maker for the contract for traditional scoring rules while distorting the information she obtains. Although there is no increase in the cost for self-financing competitive mechanisms, in that case the surplus comes from the players who are not cooperating, reducing their incentive to participate.

In situations where a decision maker can identify a particular forecaster in order to weight his report based on his reputation or other characteristics, his report can be misleading when he is acting as a coalition member rather than stating his true beliefs. One significant exception, when the information is not distorted by cooperation, is when the decision maker will use only the weighted average of the forecasts and the scoring rule is quadratic, encouraging coalitions to report their weighted average forecast.

In performing our analysis of cooperating players, we have focused on arbitrage rather than maximal expected payment. For example, if the coalition is managed by an intermediary maximizing her expected payment, she should make an identical report for all of the coalition's wagers using her personal forecast. We consider it more prudent to demonstrate how she can make a guaranteed profit without taking any risk or possessing any expertise.

Finally, while we have examined the surplus from cooperation for a coalition, we have not discussed how coalition members should behave inside a coalition. Three related issues are what coalition members will report to each other, what they will report to the decision maker, and how they will share the surplus from cooperation. We have studied this problem assuming that players can act strategically within the coalition, seeking to maximize their expected payments in the context of the coalition's decision process (Chun 2011). We assume that

- players will be paid according to a quadratic scoring rule;
- coalition members report to the decision maker the weighted average of the reports they shared within the coalition;
- coalition members distribute the surplus from cooperation proportional to their weights; and
- each coalition member believes that the expectation of the average of the other members' true probabilities is the same as his own probability,

$$\mathbb{E}_{\mathbf{p}_i}\left[\sum_{k\in\mathcal{C}}\frac{w_k}{w_\mathcal{C}}\mathbf{p}_k\right] = \mathbf{p}_i, \text{ for all } i \in \mathcal{C}.$$

These are reasonable for players who accept the "wisdom of crowds." Under these assumptions, because each coalition member wants to report truthfully within the coalition assuming that the others do so, it is a Bayesian Nash equilibrium for every coalition member to report truthfully within the coalition.

### Acknowledgements

We thank the anonymous referees and our colleagues for their suggestions, and the Samsung Foundation for four years of funding support.

## References

[1] Agrawal, S., Delage, E., Peters, M., Wang, Z., Ye, Y. (2009) A Unified Framework for Dynamic Pari-Mutuel Information Market Design, *Proceedings of the ACM Conference on Electronic Commerce (EC)*.

[2] Berg, JE and Rietz, TA (2003), Prediction Markets as Decision Support Systems, *Information Systems Frontiers* 5(1), 79-93.

[3] Brier, G.W. (1950), Verification of Forecasts Expressed in Terms of Probability, *Monthly Weather Review* 78, 1-3.

[4] Chun, S. (2011), Strictly Proper Mechanisms with Cooperating Players, Doctoral thesis, Stanford University. *(Forthcoming)*

[5] Clemen, R.T. (1989), Combining Forecasts: A review and annotated bibliography, *International Journal of Forecasting* 5, 559-583.

[6] Clemen, R.T. and Winkler, R.L. (1999), Combining Probability Distributions from Experts in Risk Analysis, *Risk Analysis* 19(2), 187-203.

[7] Clemen, R.T. (2002), Incentive Contracts and Strictly Proper Scoring Rules, *Test* 11(1), 167-189.

[8] Feldman, J., Mirrokni, V., Muthukrishnan, S., Pai, M.M. (2010), Auctions with Intermediaries, *Proceedings of the ACM Conference on Electronic Commerce (EC)* 23-32.

[9] French, S. (1985), Group Consensus Probability Distributions: A Critical Survey, *Bayesian Statistics 2*, 183-202.

[10] Graham, D.A., Marshall, R.C. (1987), Collusive Bidder Behavior at Single-Object Second-Price and English Aucions, *The Journal of Political Economy*, 95(6), 1217-1239.

[11] Hanson, R. (2003) Combinatorial Information Market Design, *Information System Frontiers* 5(1), 107-119.

[12] Hoffmann, S.A., Fischbeck, P., Krupnick, A., McWilliams, M. (2007), Elicitation from Large, Heterogeneous Expert Panels: Using Multiple Uncertainty Measures to Characterize Information Quality for Decision Analysis, *Decision Analysis* 4(2), 91-109.

[13] Johnstone D.J. (2007), The Parimutuel Kelly Probability Scoring Rule, *Decision Analysis* 4(2), 66-75.

[14] Jose, V.R.R. (2009), A Characterization for the Spherical Scoring Rule, *Theory and Decision* 66(3), 263-281.

[15] Kilgour, D.M., Gerchak, Y. (2004), Elicitation of Probabilities Using Competitive Scoring Rules, *Decision Analysis* 1(2), 108-113.

[16] Lambert, N., Langford J., Wortman, J., Chen, Y., Reeves, D., Shoham, Y., Pennock, D.M. (2008), Self-Financed Wagering Mechanisms for Forecasting, *Proceedings of the ACM Conference on Electronic Commerce (EC)* 170-179.

[17] Manski, C.F. (2006), Interpreting the Predictions of Prediction Markets, *Economic Letters* 91, 425-429.

[18] McAfee, R.P., McMillan, J. (1992), Bidding Rings, *American Economic Review* 82(3), 579-599.

[19] Morris, P.A. (1977), Combining Expert Judgments: A Bayesian Approach, *Management Science* 23(7), 679-693.

[20] Pennock, D.M. (2004) A Dynamic Pari-Mutuel Market for Hedging, Wagering, and Information Aggregation, *Proceedings of the ACM Conference on Electronic Commerce (EC)*.

[21] Pennock, D.M., Lawrence, S., Giles, C.L., Nielsen, F.A. (2001) The Real Power of Artificial Markets, *Science* 291(5506), 987-988.

[22] Peters, M., So, A., Ye, Y. (2006), A Convex Parimutuel Formulation for Contingent Claim Markets, *Proceedings of the ACM Conference on Electronic Commerce (EC)*.

[23] Savage, L.J. (1971), Elicitation of Personal Probabilities and Expectations, *Journal of the American Statistical Association* 66, 783-801.

[24] Servan-Schreiber, E., Wolfers, J., Pennock, D.M., Galebach, B. (2004), Prediction Markets: Does Money Matter?, *Electronic Markets* 14(3), 243-251.

[25] Stone, M. (1961), The Opinion Pool, *Annals of Mathematical Statistics* 32, 1339-1342.

[26] Winkler, R.L. (1996), Scoring Rules and the Evaluation of Probabilities, *Test* 5(1), 1-60.

[27] Wolfers, J., Zitzewitz, E. (2004), Prediction Markets, *Journal of Economic Perspectives*, 18(2), 107-126.

[28] Wolfers, J., Zitzewitz, E. (2005), Interpreting Prediction Market Prices as Probabilities, NBER Working Paper No. 10359.

## A  Proof of Theorem 3

We need to show that any contract for a spherical scoring rule admits arbitrage with identical report $\mathbf{q}$ for any coalition $\mathcal{C}$ of size $c \geq 2$, where

$$q_j = \frac{1}{m} + \frac{Y_j - \overline{Y}}{\sqrt{m(1 - \sum_{k=1}^m (Y_k - \overline{Y})^2)}},$$

$Y_j = \sum_{i \in \mathcal{C}} \frac{w_i p_{ij}}{w_\mathcal{C} \|p_i\|}$, and $\overline{Y} = \frac{1}{m} \sum_{j=1}^m Y_j$. The surplus from cooperation is

$$b w_\mathcal{C} \left( \sqrt{\frac{1 - \sum_{j=1}^m (Y_j - \overline{Y})^2}{m}} - \overline{Y} \right)$$

regardless of the observed outcome.

If all members of the coalition agree about $E$ then $\mathbf{q} = \mathbf{p}_i$ for all $i \in \mathcal{C}$. For the rest of this proof we will assume that the members disagree about $E$.

We will show that there is positive surplus from cooperation and that $\mathbf{q} \in \Delta_m$. First, we find an expression for $\|\mathbf{q}\|$.

$$
\begin{aligned}
\|\mathbf{q}\|^2 &= \sum_{j=1}^m q_j^2 \\
&= \sum_{j=1}^m \frac{1}{m^2} + \sum_{j=1}^m \frac{2(Y_j - \overline{Y})}{m\sqrt{m(1 - \sum_{j=1}^m (Y_j - \overline{Y})^2)}} \\
&\quad + \sum_{j=1}^m \frac{(Y_j - \overline{Y})^2}{m(1 - \sum_{j=1}^m (Y_j - \overline{Y})^2)} \\
&= \frac{1}{m} + \frac{\sum_{j=1}^m (Y_j - \overline{Y})^2}{m(1 - \sum_{j=1}^m (Y_j - \overline{Y})^2)} \\
&\quad (because \sum_{j=1}^m (Y_j - \overline{Y}) = 0) \\
&= \frac{1}{m(1 - \sum_{j=1}^m (Y_j - \overline{Y})^2)}
\end{aligned}
$$

The surplus from cooperation by reporting $\mathbf{q}$ is

$$b \sum_{i \in \mathcal{C}} \frac{w_i q_j}{\|\mathbf{q}\|} - b \sum_{i \in \mathcal{C}} \frac{w_i p_{ij}}{\|\mathbf{p}_i\|}$$

$$= b \frac{w_{\mathcal{C}}}{m \|\mathbf{q}\|} + b w_{\mathcal{C}} \overline{Y}_j - b w_{\mathcal{C}} \overline{Y} - b \sum_{i \in \mathcal{C}} \frac{w_i p_{ij}}{\|\mathbf{p}_i\|}$$

$$= b \frac{w_{\mathcal{C}}}{m \|\mathbf{q}\|} + b \sum_{i \in \mathcal{C}} \frac{w_i p_{ij}}{\|\mathbf{p}_i\|} - b w_{\mathcal{C}} \overline{Y} - b \sum_{i \in \mathcal{C}} \frac{w_i p_{ij}}{\|\mathbf{p}_i\|}$$

$$= b w_{\mathcal{C}} \left( \frac{1}{m \|\mathbf{q}\|} - \overline{Y} \right)$$

$$= b w_{\mathcal{C}} \left( \sqrt{\frac{1 - \sum_{j=1}^{m} (Y_j - \overline{Y})^2}{m}} - \overline{Y} \right),$$

regardless of the observed outcome $E_j$. Next, we prove that this surplus is strictly positive if and only if $\sum_{j=1}^{m} Y_j^2 < 1$.

$$b w_{\mathcal{C}} \left( \sqrt{\frac{1 - \sum_{j=1}^{m} (Y_j - \overline{Y})^2}{m}} - \overline{Y} \right) > 0$$

$$\Leftrightarrow \frac{1 - \sum_{j=1}^{m} (Y_j - \overline{Y})^2}{m} > \overline{Y}^2$$

$$\Leftrightarrow 1 - \sum_{j=1}^{m} (Y_j - \overline{Y})^2 > m \overline{Y}^2$$

$$\Leftrightarrow 1 - \sum_{j=1}^{m} Y_j^2 + m \overline{Y}^2 > m \overline{Y}^2$$

$$\Leftrightarrow 1 > \sum_{j=1}^{m} Y_j^2$$

$$\sum_{j=1}^{m} Y_j^2$$

$$= \sum_{j=1}^{m} \left[ \sum_{i \in \mathcal{C}} \frac{w_i p_{ij}}{w_{\mathcal{C}} \|\mathbf{p}_i\|} \right]^2$$

$$= \sum_{j=1}^{m} \left[ \sum_{i \in \mathcal{C}} \frac{w_i^2 p_{ij}^2}{w_{\mathcal{C}}^2 \|\mathbf{p}_i\|^2} + 2 \sum_{\substack{i < k \\ i \in \mathcal{C}, k \in \mathcal{C}}} \frac{w_i w_k p_{ij} p_{kj}}{w_{\mathcal{C}}^2 \|\mathbf{p}_i\| \|\mathbf{p}_k\|} \right]$$

By Cauchy-Schwarz inequality,

$$\left( \sum_{j=1}^{m} p_{ij} p_{kj} \right)^2 < \sum_{j=1}^{m} p_{ij}^2 \sum_{j=1}^{m} p_{kj}^2 = \|\mathbf{p}_i\|^2 \|\mathbf{p}_k\|^2$$

$$\Leftrightarrow \sum_{j=1}^{m} p_{ij} p_{kj} < \|\mathbf{p}_i\| \|\mathbf{p}_k\|$$

where the inequalities are strict because the coalition members disagree about $E$. Therefore,

$$\sum_{j=1}^{m} Y_j^2$$

$$= \sum_{i \in \mathcal{C}} \sum_{j=1}^{m} \frac{w_i^2 p_{ij}^2}{w_{\mathcal{C}}^2 \|\mathbf{p}_i\|^2} + 2 \sum_{\substack{i < k \\ i \in \mathcal{C}, k \in \mathcal{C}}} \sum_{j=1}^{m} \frac{w_i w_k p_{ij} p_{kj}}{w_{\mathcal{C}}^2 \|\mathbf{p}_i\| \|\mathbf{p}_k\|}$$

$$< \frac{w_i^2}{w_{\mathcal{C}}^2} + 2 \sum_{\substack{i < k \\ i \in \mathcal{C}, k \in \mathcal{C}}} \frac{w_i w_k \|\mathbf{p}_i\| \|\mathbf{p}_k\|}{w_{\mathcal{C}}^2 \|\mathbf{p}_i\| \|\mathbf{p}_k\|}$$

$$= \left( \sum_{i \in \mathcal{C}} \frac{w_i}{w_{\mathcal{C}}} \right)^2$$

$$= 1$$

We have shown that the surplus from cooperation is positive for all states of $E_j$.

Finally, we must prove that $\mathbf{q}$ lies in the simplex $\Delta_m$. Iit suffices to show that $\sum_{j=1}^{m} q_j = 1$ and $q_j \geq 0$, for all $j = 1, ..., m$.

$$\sum_{j=1}^{m} q_j = \sum_{j=1}^{m} \|\mathbf{q}\| \left[ \frac{1}{m \|\mathbf{q}\|} + Y_j - \overline{Y} \right]$$

$$= \|\mathbf{q}\| \left[ \frac{1}{\|\mathbf{q}\|} + \sum_{j=1}^{m} Y_j - m \overline{Y} \right]$$

$$= 1.$$

We have shown that $\frac{1}{m \|\mathbf{q}\|} - \overline{Y} > 0$, so

$$q_j = \|\mathbf{q}\| \left[ \frac{1}{m \|\mathbf{q}\|} + Y_j - \overline{Y} \right] > 0.$$